\documentstyle[11pt]{article}

\begin{document}
\setcounter{page}{1}
\renewcommand{\thefootnote}{\fnsymbol{footnote}}
\pagestyle{plain} \vspace{1cm}
\begin{center}
\Large{\bf Black Hole Production in the Presence of a Maximal Momentum in Horizon Wave Function Formalism}\\
\small \vspace{1cm}{\bf S. Saghafi $^{a,}$\footnote{s.saghafi@stu.umz.ac.ir}},\quad {\bf K. Nozari$^{a,b,}$\footnote{knozari@umz.ac.ir}},
\quad {\bf A. D. Kamali$^{c,}$\footnote{kamali.ata@gmail.com}}\\
\vspace{0.25cm}
$^{a}$Department of Physics, Faculty of Basic Sciences,
University of Mazandaran,\\
P. O. Box 47416-95447, Babolsar, IRAN\\
\vspace{0.25cm}
$^{b}$ Research Institute for Astronomy and Astrophysics of Maragha (RIAAM),\\
P. O. Box 55134-441, Maragha, Iran\\
\vspace{0.25cm}
$^{c}$Department of Physics, Science and Research Branch,\\
Islamic Azad University, Tehran, Iran\\

\end{center}
\vspace{1.5cm}
\begin{abstract}

We study the Horizon Wave Function (HWF) description of a generalized uncertainty
principle (GUP) black hole in the presence of two natural cutoffs as a minimal length and a maximal momentum. This is motivated by a metric which allows  the existence of sub-Planckian black holes, where the black hole
mass $m$ is replaced by $M=m\Big(1+\frac{\beta^{2}}{2}\frac{M_{pl}^{2}}{m^{2}}-\beta\frac{M_{pl}}{m}\Big)$. Considering a wave-packet
with a Gaussian profile, we evaluate the HWF and the probability that the source might be a (quantum) black hole. By decreasing the free parameter the general form of probability distribution, ${\mathcal{P}}_{BH}$, is preserved , but this resulted in reducing the probability for the particle to be a black hole accordingly. The probability for the particle to be a black hole grows
when the mass is increasing slowly for larger positive $\beta$, and for a minimum mass
value it reaches to $0$. In effect, for larger $\beta$ the magnitude of $M$ and $r_{H}$ increases, matching with our intuition that either the particle ought to be more localized or more massive
to be a black hole. The scenario undergoes a change for some values of $\beta$ significantly, where there is a minimum in
${\mathcal{P}}_{BH}$ , so this expresses that every particle can have some probability of decaying
to a black hole. In addition, for sufficiently large $\beta$ we find that every particle could be fundamentally a
quantum black hole.
\\
{\bf PACS}: 04.70.-s, 04.70.Dy\\
{\bf Key Words}: Quantum Gravity, Generalized Uncertainty Principle, Black Holes, Horizon Wave Function
\end{abstract}

\section{Introduction}

Our present understanding of the Universe relies on two theories: the theory of
General Relativity which is applicable at very large scales, and Quantum Mechanics
which provides a very good description of the microscopic universe. While these two
theories work very well in their own regimes, the regime in which they collide
is near the Planck scale, when quantum mechanical wave-packets can turn into Black Holes.
Another interesting problem is the description of the gravitational collapse which
again leads to the formation of black holes (firstly investigated in the seminal papers of Oppenheimer \emph{et al.} [1,2]).
A lot of work has been done on this subject, but a good description of the physics of such processes is still challenging.
A vast amount of research has been done on this subject (see, e.g. Ref. [3]), but many
conceptual problems remain unsolved, such as accounting for the quantum mechanical nature of
collapsing matter.\\

Until now, the only unanimously accepted idea is that gravitation becomes
important whenever a large enough amount of matter is compacted within a sufficiently small volume. K. Thorne was the first one who formulated this idea in the hoop conjecture [4],
which remarks that a BH forms when two objects which collide together, collapse within their black
disk. presuming the final structure is (approximately) spherically symmetric,
this happens when the collapsing objects occupy a sphere with radius $r$ and this radius is smaller than the Schwarzschild radius,
\begin{equation}
r\leq R_{H}\equiv 2\ell_{pl}\frac{E}{m_{pl}}
\end{equation}
where the Planck length is denoted by $\ell_{pl}$ (defined as $\ell_{pl}=\sqrt{\frac{G\hbar}{c^{3}}}$) and $m_{pl}$ represents the Planck mass.\\

One can find many attempts at quantizing BH metrics in the literature, which
focus on the purely gravitational degrees of freedom, and result in a description of
the horizon which is unrelated to the matter state that sourced the geometry to begin with [5]. The approach discussed in this paper, is
one of the approaches in comprehending the essence of quantum black holes by considering the quantum mechanical conditions for their creation according to their horizon wave function. In this formalism which is called the \emph{horizon wave function} (HWF) approach, one can suppose that the black hole is a quantum mechanical particle, so a spatial wave function can be attributed to this black hole which is placed in its classical event horizon [6,7]. If these particles can be created in high energy collisions, so the probability for creating a black hole can be assessed by the corresponding collisions probability. Furthermore, the HWF has been studied for understanding the quantum nature of black hole thermodynamics [6-8]. Also in [7], HWF has been applied to the lower and higher dimensional models and the end of black hole evaporation in a lower dimensional model is studied in [9]. The most appealing feature of the HWF approach is the aspect of generalized uncertainty principle in which the wave and particle gravitational length scales will affect the quantum uncertainties. In fact it is shown that as a property of string theory, generalized uncertainty principle (GUP) is one of the proposed outcomes of the final quantum gravity theory which essentially has no dependence on models. That is, although the functional form of GUP (and therefore modified dispersion relations) are model dependent, existence and the very nature of these uncertainties are model independent. These theories are loop quantum gravity [10], non commutative geometry [11] and other minimal length scale scenarios such as the doubly special relativity. Even though most of minimal length scale approaches consider a minimal limit for black hole mass, recently in Ref. [12] the authors found a new generalized uncertainty principle that modifies the Schwarzschild metric and admits the existence of sub-Planckian black holes. In this paper we consider the effect of adding an additional term to this GUP because of the existence of maximal momentum. This extra feature is a result of the doubly special relativity theories [13]. As the HWF approach can predict the probability of black hole formation for arbitrary masses and also the origin of this horizon wave function should be limited due to the uncertainty principle, we are looking for the information that how the probability of black hole formation can be affected by the generalized uncertainty principle in the black hole metric. We focus also on the role of maximal momentum on the probability distribution, ${\mathcal{P}}_{BH}$.

\section{Black holes and Generalized Uncertainty Principle}
  When one approaches very small length scales (towards the Planck length), the standard Heisenberg's uncertainty relation gets modified to, for instance, the following generalized uncertainty principle [14,15,16]
$$\Delta x \Delta p \geq \hbar \Big[1-\beta \Delta p + \beta^{2} (\Delta p)^{2}\Big]$$
which gives
\begin{eqnarray}
\Delta x\geq \hbar \Big[\frac{1}{\Delta p}-\beta +\beta^{2}\Delta p\Big]\,.
\end{eqnarray}
In this relation $\beta=\beta_{0}\frac{\ell_{pl}^{2}}{\hbar^{2}}$ and $\beta_{0}$ is a dimensionless constant parameter. This GUP admits two natural cutoffs as a minimal length and a maximal momentum and can cause a similar relation for momentum and mass in the domain of characteristic length of the system.
In Ref.~[12] authors discussed that the black hole uncertainty principle correspondence propose that black holes with mass smaller than the Planck scale but radius of order the Compton scale instead of the Schwarzschild scale, could exist. They introduce a modified self-dual Schwarzschild like metric that though in the large mass limit it remains Schwarzschild , reproduces acceptable aspects of a variety of several models in sub-Planckian limit. They considered a Generalized Uncertainty Principle which admits the existence of a minimal length. In this paper we also considered the effect of maximal momentum. The main task of this case is that the generalized uncertainty principle is situated in the very notion of the geometry of spacetime, so the metric which is used for describing this spacetime must have a specific relation for mass. In the limit of large masses, $M>M_{pl}$, where quantum effects are insignificant, the Schwarzschild solution can be recovered. In this case the black hole mass can be defined in terms of energy-momentum tensor. When $M<M_{pl}$, the exact meaning of mass parameter becomes ambiguous [12]. One can consider both a black hole and particle. As the size of horizon of such sub-Planckian black holes is smaller than the Planck length, relativistic description becomes deficient. Hence, in this case there would be a particle of mass $M\sim\frac{\hbar}{\lambda_{c}}$ where $\lambda_{c}$ is the Compton wavelength. Even this mass can be represented as a kind of Komar mass as
\begin{equation}
M\equiv\int_{\Sigma}d^{3}x\,\gamma\, n_{\mu}\kappa_{\nu}\,T^{\mu\nu}\cong -4\pi\int_{0}^{c}dr\,r^{2}\,T_{0}^{0}\,,
\end{equation}
in which $\gamma$ is the deformed spatial determinant of $\gamma^{ij}$, $T^{\mu\nu}$ is energy-momentum tensor and $T^{0}_{0}$ measures the energy density on the length scale $\lambda_{c}$.
In the absence of a full quantum gravity theory, the exact form of energy-momentum tensor is ambiguous, but one can consider that $T_{0}^{0}$ shows the quantum mechanical distribution of matter [12]. So the exact definition of mass both contains the large scale mass (ADM) and also the mass of particles in small scale. As in [12] the dual role of $M$ has been inspired in the GUP ,  now we can suggest that the Arnowitt-Deser-Misner (ADM) mass which coincides with the Komar energy in the stationary case, should be

\begin{equation}
M_{ADM}=M\Big(1+\frac{\beta^2}{2}\frac{M_{pl}^2}{M^2}-\beta \frac{M_{pl}}{M}\Big)
\end{equation}

On the basis of the above discussion and according to what is done in [12], the quantum deformed form of the Schwarzschild metric is as follows

\begin{equation}
ds^{2}=F(r)dt^{2}-F^{-1}(r)dr^{2}-r^{2}d\Omega^{2}\,,
\end{equation}
where
\begin{equation}
F(r)\equiv 1-\frac{2}{m_{pl}^2}\frac{M}{r}\Big(1+\frac{\beta^{2}}{2}\frac{m_{pl}^2}{M^{2}}-\beta\frac{m_{pl}}{M}\Big)\,.
\end{equation}
In fact, this metric enfolds all the characteristics of the Schwarzschild spacetime as this modified term is a coordinate-independent relation. In this metric the event horizon can be written as
\begin{equation}\label{1}
r_{H}=\frac{M+(M+\beta m_{pl})^{2}}{M m_{pl}^{2}}
\end{equation}
which leads to
\begin{equation}
M\gg m_{pl} \longrightarrow r_{H}\simeq \frac{2M}{m_{pl}^{2}}\,,
\end{equation}
\begin{equation}
M=m_{pl} \longrightarrow r_{H}\simeq \frac{1+m_{pl}(1+\beta)^{2}}{m_{pl}^{2}}\,,
\end{equation}
\begin{equation}
M\ll m_{pl} \longrightarrow r_{H}\simeq \frac{\beta}{M}\,,
\end{equation}
for black holes which have the mass of super-Planckian, Planckian and sub-Planckian respectively. Figure 1 shows the behavior of event horizon as a function of mass for different values of $\beta$.\\

In Ref. [12] it is shown that although the singularity is not removed, it can never be reached. This can be proved by inverting the relation (\ref{1}) according to two masses relating to the given horizon which leads to a minimal horizon radius as
\begin{equation}
r_{min}=\frac{2\beta}{m_{pl}}\,,
\end{equation}
 and also a minimum mass as
\begin{equation}
M_{min}=\beta m_{pl}\,.
\end{equation}

\begin{figure}[htp]
\begin{center}\includegraphics{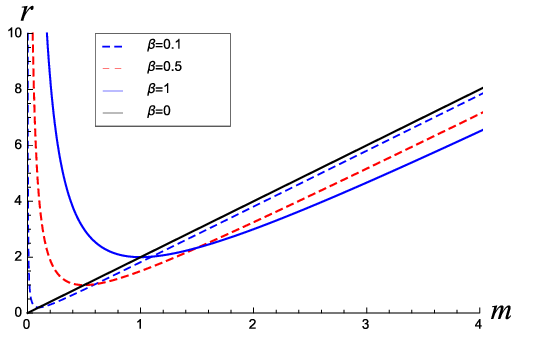} \vspace{4 cm}
\end{center}
\caption{\small {Event horizon as a function of mass. }}
\end{figure}

\section{GUP Black Holes in Horizon Wave Function Formalism}

As it was mentioned in Refs.~[6,7], the horizon is a classical concept in general relativity and it cannot be defined clearly when the source is not described by classical physics. One can consider a horizon wave function for any localized quantum mechanical wave-function that can lead to compute the probability of finding a horizon of a given radius centralized around the source. So a probability can be associated to each quantum particle that is a black hole and also to the existence of minimum black hole mass. This agrees with the results of the hoop conjecture [17] and the Heisenberg Uncertainty Principle: the black hole can form whenever the impact parameter $b$ of two objects which are colliding is shorter than the Schwarzschild radius of the system, it means:

\begin{equation}\label{hop}
b\leq 2\ell_{pl}\frac{E}{m_{pl}}\equiv r_{H}
\end{equation}

where $E$ is total energy in the centre of mass frame. This conjecture has been checked in
a variety of situations. For understanding a classical horizon in a spherically
symmetric space-time,a general spherically symmetric metric $g_{\mu\nu}$ can be written as

\begin{equation}
ds^{2}=g_{ij}dx^{i}dx^{j}+r^{2}(x^{i})(d\theta^{2}+sin^{2}\theta d\phi^{2})
\end{equation}

By considering this classical metric, we can find out exactly the location of a trapping horizon, a surface where the escape
velocity is equal to the speed of light, as [18] :

\begin{equation}\label{tr}
0=g_{ij}\nabla_{i}r \nabla_{j}r=1-\frac{2\ell_{pl}(\frac{m}{m_{pl}})}{r}.
\end{equation}

where $\nabla_{i}r $ is defined as the covector perpendicular to surfaces of constant area $A=4\pi r^{2}$. The
function $M =\frac{\ell_{pl}m}{m_{pl}}$ is the active gravitational (or Misner-Sharp) mass, which shows
the total energy which is surrounded by a sphere of radius $r$. It is obvious that in general
following the dynamics of a given matter distribution and verifying the existence of surfaces satisfying Eq.\ref{tr} is so complex, but an horizon can be found if there exists values of r such that
$r_{H}=2M(t,r)>r$, which is reexpressing of the hoop conjecture mathematically (\ref{hop}).\\

According to what is done in [6], now we consider a point-like  mass $m$ which is also a spin-less particle, and it's Schwarzschild radius
is given by $r_H$ in Eq. (\ref{hop}) with $E=m$. For such a particle, the Heisenberg principle of
quantum mechanics introduces an uncertainty in its spatial localisation, typically of the
order of the Compton-de Broglie length

\begin{equation}
\lambda_{m}\simeq \ell_{pl} \frac{m_{pl}}{m}.
\end{equation}

As quantum physics is a more processed description of reality, the conflict of the two
lengths, $r_H$ and $\lambda_{m}$, points that it only makes sense if:

\begin{equation}
r_{H} \geq \lambda_{m}\Rightarrow m \geq m_{pl}.
\end{equation}

Now the spread in localization can be represented by the wavefunction

\begin{equation}
\mid \Psi_{S}\rangle =\sum_{E} C(E)\mid \Psi_{E}\rangle,
\end{equation}

As usual, the sum over the variable $E$ represents the decomposition on the spectrum of the
Hamiltonian,

\begin{equation}
\hat{H} \mid \Psi_{E}\rangle=E\mid \Psi_{E}\rangle
\end{equation}

Once the energy spectrum is known, we can use (\ref{hop}) to get

\begin{equation}
E=m_{pl}\frac{r_{H}}{2\ell{p}}
\end{equation}

Now the HWF can be defined as

\begin{equation}
\Psi_{H}(r_{H})=C(\frac{m_{pl}r_H}{2\ell_{pl}})
\end{equation}

and can be normalized as
\begin{equation}\label{2}
\langle\psi_{H}\mid \phi_{H}\rangle=4\pi \int_{0}^{\infty} \psi^{*}_{H}(r_{H})\phi_{H}(r_{H})r_{H}^{2}dr_{H}\,.
\end{equation}
The concept of normalized HWF $|\psi_{H}\rangle$ is that it is the probability that an observer measures the particle at the quantum state $|\psi_{S}\rangle$ and associates it to an event horizon with the radius $r=r_{H}$. As a result, the defined classical horizon is replaced by the expectation value of operator $r_{H}$. The probability of the gravitational source to be a black hole is that it should be located totally in its horizon
\begin{equation}\label{bh}
{\cal P}_{BH}=\int_{0}^{\infty} P_{<}(r<r_{H})dr_{H}\,,
\end{equation}
in which the probability density
\begin{equation}
P_{<}(r<r_{H})=P_{S}(r<r_{H})P_{H}(r_{H}),
\end{equation}
is a combination of the probability that a particle be at rest in a sphere with the radius $r=r_{H}$ and also the probability that $r_{H}$ is gravitational radius. These quantities can be calculated as
\begin{eqnarray}
P_{S}(r<r_{H})=\int_{0}^{r_{H}}P_{S}(r)dr=4\pi\int_{0}^{r_{H}}\mid\psi_{S}(r)\mid^{2}r^{2}dr\,, \\
P_{H}(r_{H})=4\pi r_{H}^{2}\mid \psi_{H}(r_{H})\mid^{2}.
\end{eqnarray}

According to what is done in Refs.~[7,19,20,21,22], a massive particle at rest in the reference frame can be characterized by the following Gaussian profile
\begin{equation}
\psi_{S}(r)=\frac{e^{-\frac{r^2}{2\ell^2}}}{(\ell\sqrt \pi)^{\frac{3}{2}}}\,.
\end{equation}
Now with the inspiration by twofold role of mass $m$ in the generalized uncertainty principle, one can investigate the existence of sub-Planckian black holes meaning that they are both particle and black hole. In this framework the usual $m$ is replaced by the modified $M$ of the generalized uncertainty principle as

\begin{equation}
M=m\Big(1+\beta^{2}\frac{m_{p}^{2}}{m^{2}}-\beta\frac{m_{p}}{m}\Big)\,.
\end{equation}

In the next step we consider a case that the length $\ell$ is related to the uncertainty in the size of the particle. It can be equal to the Compton length as

\begin{equation}
\ell=\lambda_{m}\simeq\frac{\ell_{p}m_{p}}{M}.
\end{equation}

Remembering that these analysis are appropriate for independent $\ell$ and $m$, this case is related to the maximum localization for the gravitational source as one expects,  $\ell\geq\lambda_{m}$.\\

By taking the Fourier Transform, the corresponding wavefunction in momentum space gives

\begin{equation}
\tilde{\psi}_{S}(p)=\frac{e^{-\frac{p^2}{2\Delta}}}{(\Delta\sqrt{\pi})^\frac{3}{2}}
\end{equation}

Assuming the relativistic mass-shell equation in flat space-time to account for high-energy
particle collisions, we relate the momentum $p$ to the total energy $E$

\begin{equation}
E^{2}=p^2+M^2
\end{equation}

From the Schwarzschild relation (\ref{1}) and by maintaining the normalization according to (\ref{2}), we derive the HWF as

\begin{equation}
\psi_H(r_H)\simeq \Theta(r_H-r_{min})\exp\left[-\frac{\ell^2}{8\,\ell_p^4}
\left(r_H^2-r_{min}^2\right)\right]\,,
\end{equation}
in which $r_{H}=\frac{2M}{m_{p}^{2}}$ and the Heaviside step function results in from of the fact that $E\geq M$. \\
Now the density probability can be explicitly computed as follows

\begin{eqnarray}
P_<=\frac{\ell^3r_H^2e^{-\frac{\ell^2\,r_H^2}{4}}}{2\sqrt{\pi}}
\left[{\rm Erf}\left(\frac{r_H}{\ell}\right)
-\frac{2r_He^{-\frac{r_H^2}{\ell^2}}}{\sqrt{\pi}\ell}\right]
\end{eqnarray}

and

\begin{eqnarray}
{P_ < }= \frac{{2{\rm{Erf}}\left( {2{m^2}{{\left( {1 + \frac{{{\beta ^2}}}{{{m^2}}} - \frac{\beta }{m}} \right)}^2}} \right)}}{{e\sqrt \pi  m\left( {1 + \frac{{{\beta ^2}}}{{{m^2}}} - \frac{\beta }{m}} \right)}}-\frac{8}{\pi }\exp \left[ { - 4{m^4}{{\left( {1 + \frac{{{\beta ^2}}}{{{m^2}}} - \frac{\beta }{m}} \right)}^4} - 1} \right],
\end{eqnarray}
from which by integrating the density from $r_{H}$ to infinity, the probability (\ref{bh}) for a particle to be a black hole is obtained as
\begin{eqnarray}
{\cal{P}}_{BH}(\ell ) = \frac{2}{\pi }\left[ {\arctan \left( {{\mkern 1mu} \frac{2}{{{\ell ^2}}}} \right) - \frac{{2{\mkern 1mu} {\ell ^2}{\mkern 1mu} ({\ell ^4} - 4)}}{{{\mkern 1mu} ({\ell ^4} + 4)}}} \right]\,.
\end{eqnarray}
By writing ${\cal P}_{BH}$ as a function of $m$, this relation can be rephrased as follows
\begin{eqnarray}
&&{{\cal P}_{{\rm{BH}}}}(m) = \frac{2}{\pi }\left[ {\arctan \left( {2 {m^2}{{\left( {1 + \frac{{{\beta ^2}}}{{{m^2}}} - \frac{\beta }{m}} \right)}^2}} \right)} \right]\nonumber\\
 &+&\frac{{ 16 - \frac{4}{{{m^4}{{\left( {1 + \frac{{{\beta ^2}}}{{{m^2}}} - \frac{\beta }{m}} \right)}^4}}}}}{{4\pi {m^2}{{\left( {1 + \frac{{{\beta ^2}}}{{{m^2}}} - \frac{\beta }{m}} \right)}^2}  - \frac{1}{{{m^2}{{\left( {1 + \frac{{{\beta ^2}}}{{{m^2}}} - \frac{\beta }{m}} \right)}^2}}}}}\,.
\end{eqnarray}

\begin{figure}[htp]
\begin{center}\includegraphics{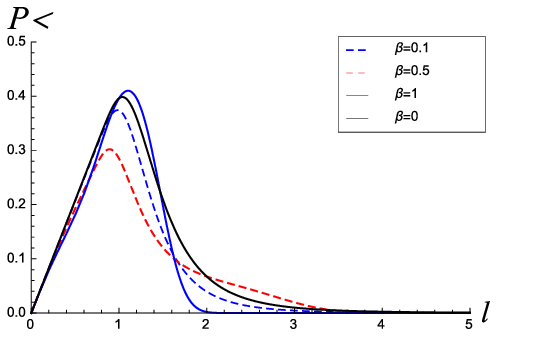} \vspace{4cm}
\end{center}
\caption{\small {Probability density as a function of Gaussian width $\ell$ for some values of $\beta$. }}
\end{figure}

\begin{figure}[htp]
\begin{center}\includegraphics{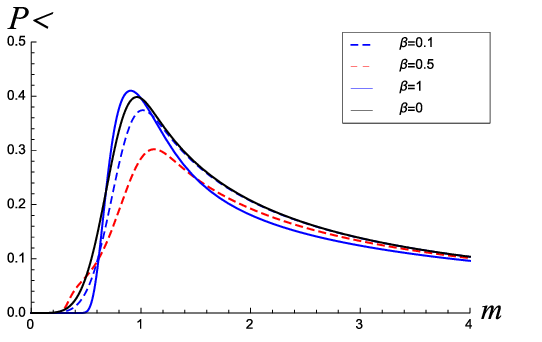} \vspace{3.5cm}
\end{center}
\caption{\small {Probability density as function of mass for some values of $\beta$. }}
\end{figure}

\begin{figure}[htp]
\begin{center}\includegraphics{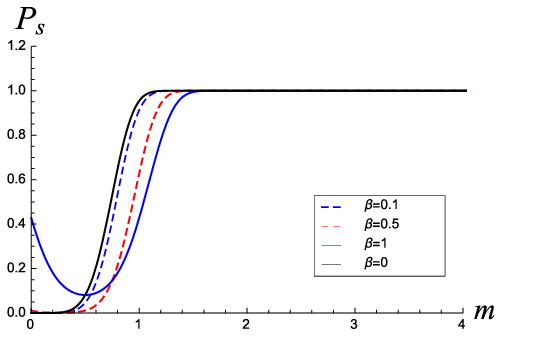} \vspace{5cm}
\end{center}
\caption{\small {Probability that a particle to be a black hole versus its mass for some values of $\beta$. }}
\end{figure}

Figures ($2$) and ($3$) show the probability densities as a function of the size and mass of the particle for different values of $\beta$, that is, the probability for the particle to be inside its own horizon. Also the probability that the particle is a black hole is plotted as a function of mass $m$ in Fig. ($4$). This scenario has some limitation because of the cutting mass and as it was discussed previously, this cutting mass is related to the minimum radius $r_{min}$ which is $M_{_{r_{min}}}=\beta m_{_P}$. The masses which are smaller than this cutoff are just used in numerical calculation. It should be mentioned that for very small $\beta$ the probability that a particle to be a black hole exactly resembles the standard Schwarzschild metric [7]. This result is in agreement with the modified theory here, which for $\beta=0$ , $M\rightarrow m$ and the horizon radius equals to the classical Schwarzschild radius.\\
On the other hand, for some values of $\beta$ there would be a minimum in their probability according to the Fig. 4; it means that for any amount of particle's mass there would be a certain probability to be a black hole (when $\beta=1$ this case is more obvious). Also, for enough large $\beta$, everything can be a black hole, in accommodation with the fact that the GUP introduced here does'nt exert any minimal mass directly. So the effect of increasing the free parameter $\beta$ would be gathering larger masses in a larger horizon radius. Therefore, it is more probable that a particle to be a black hole. An important property in this regard is that $m\geq M_{pl}$ will result in $P_{BH}\simeq 1$. Another issue that is important to note is that the existence of a maximal momentum encoded in the term $-\beta\frac{m_{p}}{m}$ in equation (18), reduces the value of $M$ for positive $\beta$. This reduction of $M$ leads to a less probability of forming black hole. So, maximal momentum cutoff reduces the black hole production probability in essence.

\section{Summary and Conclusion}

In this paper the probability of black hole formation in a gravitational field and within the horizon wave function approach is studied in the presence of natural cutoffs as a minimal length and maximal momentum. Our focus is mainly on the role of maximal momentum in this setup. For this goal, the gravitational source is considered to be immersed in a modified metric, the modification of which are coming from the GUP consisting a minimal measurable length and a maximal momentum. By considering a wave packet with a Gaussian profile, the corresponding horizon wave function is computed and the probability that the gravitational source is a black hole (a quantum black hole), i.e. situating in its horizon radius, is derived. We treated the problem by some numerical analysis on the probability as a function of the mass for different values of $\beta$. First of all we observed that when increasing the positive free parameter $\beta$, a minimum in ${\cal{P}}_{BH}$ would occur which means that essentially every particle can collapse into a black hole in this situation. More ever, for large $\beta$, every particle is essentially a quantum black hole because by increasing $\beta$, $M$ and $R_{H}$ would become larger and due to this fact it would be more probable for a particle to lie in its event horizon and therefore to be a black hole. An important result of inclusion of maximal momentum cutoff in our setup is that for positive beta, the resulting modified mass, $M$, reduces (by the factor $-\beta\frac{m_{p}}{m}$) which reduces probability of a particle to be a quantum black hole in this framework. It should be mentioned that even though we restrict the sub-Planckian black holes to a mass-cutoff, but the exact essence of black holes especially in extremely small masses depends on which theory of quantum gravity would be authentic finally.\\

{\bf Acknowledgement}\\
The work of K. Nozari has been supported financially by Research
Institute for Astronomy and Astrophysics of Maragha (RIAAM) under
research project number 1/6025-00.

\end{document}